\documentclass{ws-ijmpa}
\usepackage[super,compress]{cite}
\usepackage{graphicx}
 \def\lra#1{\overset{\text{\scriptsize$\leftrightarrow$}}{#1}}
\begin{document}
\markboth{John Ellis}{Where is Particle Physics Going?}

%
\catchline{}{}{}{}{}
%

\title{\bf WHERE IS PARTICLE PHYSICS GOING?}

\author{JOHN ELLIS}

\address{Theoretical Particle Physics and Cosmology Group, Physics Department, \\ KingÕs College London, London WC2R 2LS, UK; \\
Theoretical Physics Department, CERN, CH-1211 Geneva 23, Switzerland \\
John.Ellis@cern.ch}

\maketitle

\begin{history}
\received{Day Month Year}
\revised{Day Month Year}
\end{history}

\begin{abstract}
The answer to the question in the title is: in search of new physics beyond the Standard Model, for which there are many motivations, including
the likely instability of the electroweak vacuum, dark matter, the origin of matter, the masses of neutrinos,
the naturalness of the hierarchy of mass scales, cosmological inflation and the search for quantum gravity.
So far, however, there are no clear indications about the theoretical solutions to these problems, nor the experimental
strategies to resolve them. It makes sense now to prepare various projects for possible future accelerators,
so as to be ready for decisions when the physics outlook becomes clearer. 
Paraphrasing George Harrison, `` If you don't {\it yet} know where you're going,
any road {\it may} take you there."\\
~\\
{\it Contribution to the 2017 Hong Kong UST IAS Programme and Conference on High-Energy Physics.}\\
~\\
KCL-PH-TH-2017-18, CERN-TH-2017-080

\end{abstract}

\ccode{PACS numbers: 12.15.-y, 12.60.Jv, 14.80.Bn, 14.80.Ly}


\section{Introduction}	
\begin{figure}[t!]
\begin{center}
\includegraphics[height=6cm]{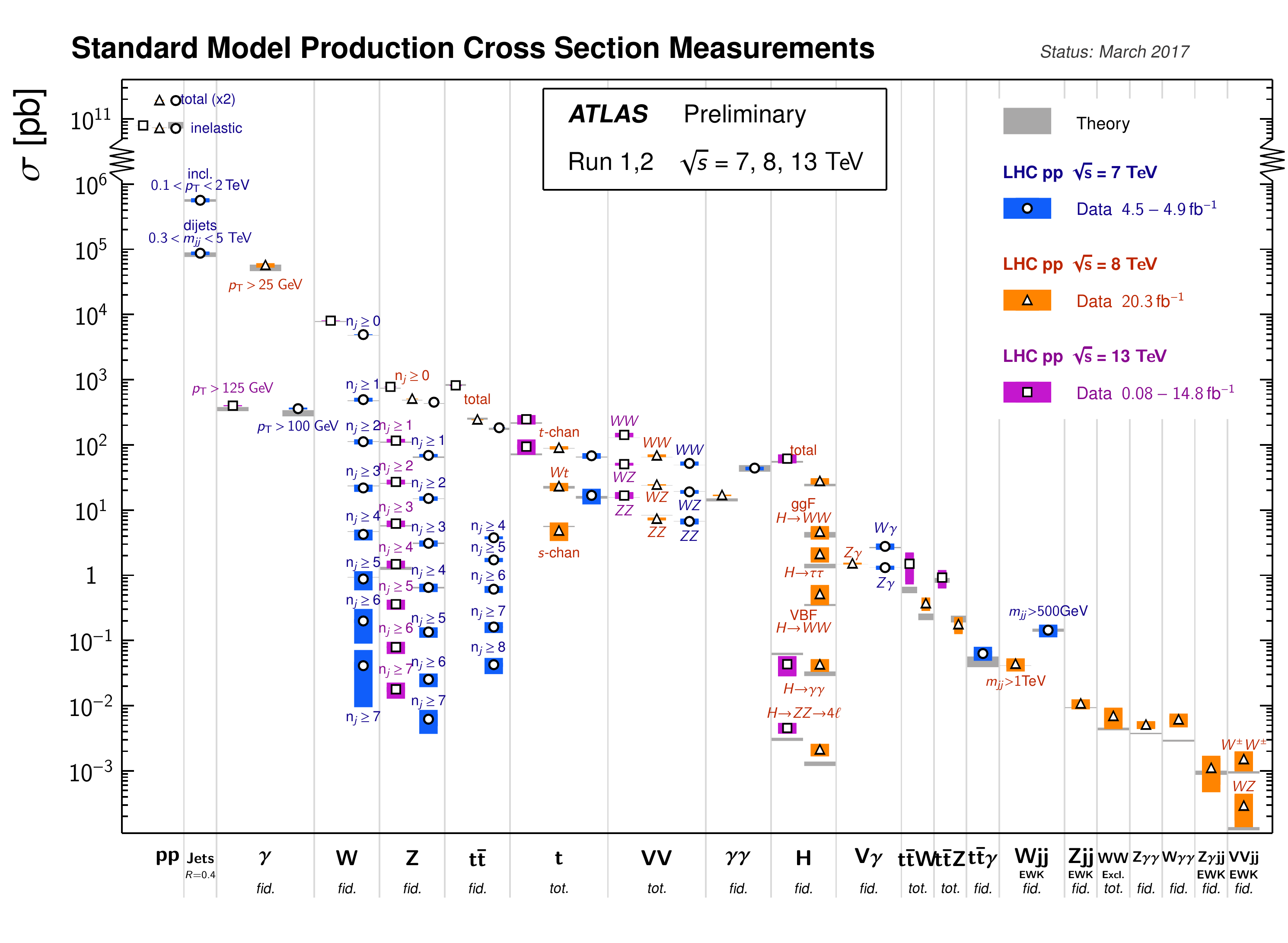}
\end{center}   
\caption{\label{fig:heaven}\it 
Many SM processes have been measured at the LHC, and have cross sections that are generally in excellent agreement with
QCD calculations~\protect\cite{ATLASSM}.
}
\end{figure}

The bedrock upon which our search for new physics beyond the Standard Model (SM) is founded
is our ability to make precise predictions within the Standard Model, notably for the LHC
experiments. The predictions of many hard higher-order perturbative QCD calculations have been confirmed,
as seen in Fig.~\ref{fig:heaven}, providing confidence in predictions for the production of the 
Higgs boson~\cite{Mistlberger}, and for the backgrounds to many searches for new physics.

\section{The Flavour Sector}

\begin{figure}[t!]
\begin{center}
\includegraphics[height=5.5cm]{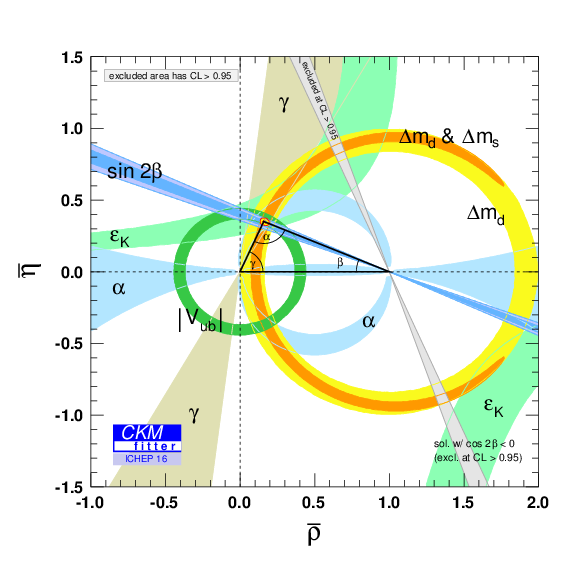}
\includegraphics[height=5.1cm]{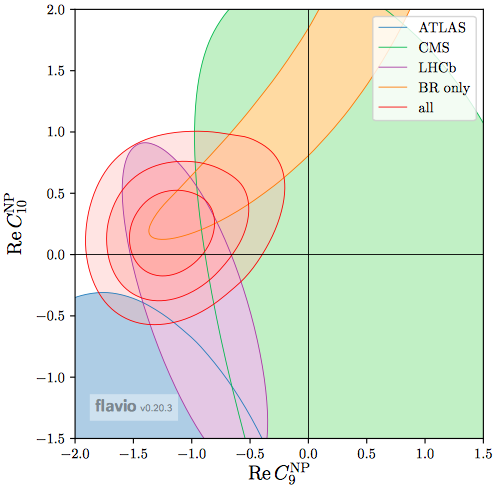} \\
\end{center}   
\caption{\label{fig:CKM}\it 
Left panel: Compilation of experimental constraints on the CKM unitarity triangle~\protect\cite{CKMfitter}.
Compilation of constraints on  possible new physics
contributions to operator coefficients~\protect\cite{Wolfgang}.}
\end{figure}

Many measurements in the flavour sector are also consistent with the predictions of the
Cabibbo-Kobayashi-Maskawa (CKM) model~\cite{CKMfitter,UTfit}, e.g., there are many consistent measurements of the
unitarity triangle, as seen in the left panel of Fig.~\ref{fig:CKM}. Historically, the angle $\gamma$ has been the least
constrained experimentally, but the LHCb Collaboration has recently published a combined
measurement~\cite{LHCbgamma} that dominates the world average and is consistent with the other unitarity
triangle measurements.

That said, there are several anomalies in the flavour sector of varying significance.
For example, there are strengthening indications of violations of $e/\mu$ lepton universality in
$B \to K e^+ e^-$ and $B \to K \mu^+ \mu^-$ decays,~\cite{Kll} and of $\tau/(\ell = e$ or $\mu)$ universality in
$B \to D^{(*)} \tau \nu$ decays~\cite{Dtaunu} - to which my attitude is `wait and see', as lepton non-universality
has held up very well so far. Much attention has been attracted to the $P_5^\prime$ angular
distribution in $B \to K^* \mu^+ \mu^-$ decay~\cite{P5prime}, which may be accompanied by an anomaly in
the $q^2$ distribution in $B \to \phi \mu^+ \mu^-$ decay, leading to the constraints on possible new physics
contributions to operator coefficients shown in the right panel of Fig.~\ref{fig:CKM}~\cite{Wolfgang}. These both appear at $q^2 \lesssim 5$~GeV$^2$,
and I do not know how seriously to take them, in view of my lack of understanding of the non-perturbative 
QCD corrections in this region. My ignorance also makes it difficult for me to judge the significance
of the apparent discrepancy between theory~\cite{CTS,Buras} and experiment for $\epsilon^\prime/\epsilon$. Finally,
a new kid on the flavour block has been the interesting search for $H \to \mu \tau$ decay~\cite{Htaumu} discussed below, though this may be
reverting towards the SM with the latest Run~2 results~\cite{CMSHtaumu2}.

\section{Higgs Physics}
\subsection{The Higgs Mass}

The most fundamental Higgs measurement is that of its mass. The combined LHC Run~1 results
of ATLAS and CMS based on $H$ decays into $\gamma \gamma$ and $Z Z^* \to 2 \ell^+ 2 \ell^-$ yielded~\cite{ATLAS+CMS}
\begin{equation}
m_H \; = \; 125.09 \pm 0.21 ({\rm stat.}) \pm 0.11 ({\rm syst.}) \, ,
\label{mHRun1}
\end{equation}
and the preliminary CMS result from Run~2 is consistent with this, with slightly smaller errors~\cite{CMSRun2}:
\begin{equation}
m_H \; = \; 125.26 \pm 0.20 ({\rm stat.}) \pm 0.08 ({\rm syst.}) \, ,
\label{mHRun2}
\end{equation}
It is noteworthy that statistical uncertainties dominate, and we can look forward to substantial
reductions in the future, determining $m_H$ at the {\it per mille}. 
Accurate knowledge of the Higgs mass is important for precision tests
of Standard Model (and other) predictions and, as discussed later, is crucial for understanding
the (in/meta)stability of the electroweak vacuum.

\subsection{Higgs Couplings}

The couplings of the Higgs boson to Standard Model particles are completely specified and,
consequently, there are definite predictions for its production processes and decay branching
ratios~\cite{LHCHXSWG}. Concretely, one expects gluon-gluon fusion to dominate over vector-boson fusion,
production in association with a vector boson and in association with a $t {\bar t}$ pair.
The dominant $H$ decay mode is predicted to be into $b {\bar b}$, with much smaller 
branching ratios for $\gamma \gamma$ and $Z Z^* \to 2 \ell^+ 2 \ell^-$.

Much progress was made in Run~1 probing these predictions~\cite{ATLAS+CMS2}, but much remains to be done.
Higgs decays to $\gamma \gamma, ZZ^*, WW^*$ and $\tau^+ \tau^-$ have been measured
in gluon-gluon fusion, and there is solid evidence for vector-boson fusion, but the associated
production mechanisms have yet to be confirmed. Moreover, there is no confirmation yet of
the expected dominant $H \to b {\bar b}$ decay mode: LHC evidence is at the level of 2.6 $\sigma$~\cite{Tevatronbbbar},
and the Tevatron experiments have reported evidence at the 2.8-$\sigma$ level. There is
indirect evidence for the expected $H t {\bar t}$ vertex via the measurements of gluon-gluon fusion
and $H \to \gamma \gamma$ decay, but no significant evidence via associated $H t {\bar t}$
or single $H t ({\bar t})$ production. Also on the agenda is the search for $H \to \mu^+ \mu^-$,
which is predicted in the SM to appear at a level close to the current experimental sensitivity.

\begin{figure}[ht]
\begin{center}
\includegraphics[width=7cm]{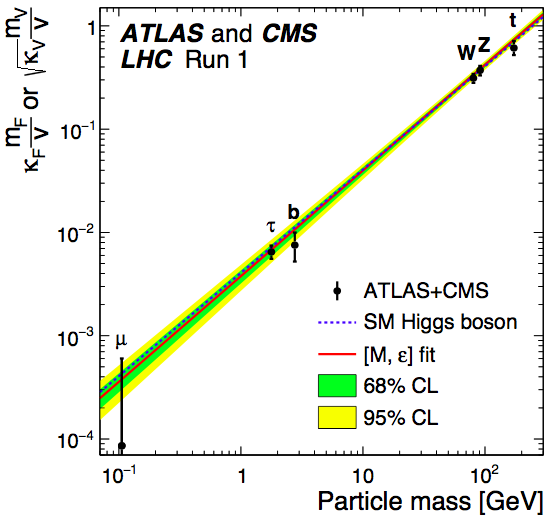}
\caption{\it A fit by the ATLAS and CMS Collaborations to a parametrization of the mass-dependence of the Higgs couplings: 
$m \to m^{(1+ \epsilon)}$~\protect\cite{ATLAS+CMS2}.
The Standard Model predictions are connected by a dotted line, the red line is the best fit,
and the green and yellow bands represent the 68 and 95\% CL fit ranges.}
\label{fig:Mepsilon}
\end{center}
\end{figure}

Fig.~\ref{fig:Mepsilon} is one way of displaying the available information on Higgs couplings~\cite{EY,ATLAS+CMS2}.
It is a characteristic prediction of the SM that the couplings to other particles should be
related to their masses, $\propto m_f$ for fermions and $\propto m_V^2$ for massive
vector bosons. The black solid line is a fit where $m \to m^{(1+ \epsilon)}$ in the couplings: we
see that the combined ATLAS and CMS data are highly consistent with the SM expectation
that $\epsilon = 0$, shown as the blue dashed line.

The couplings in Fig.~\ref{fig:Mepsilon} are all flavour-diagonal. The SM predicts that
flavour-violating Higgs couplings should be very small, but measurements of flavour-violating
processes at low energies would allow {\it either} $H \to \mu \tau$ {\it or} $H \to e \mu$ with
 branching ratio $\lesssim 10$\%, whereas the branching ratio for $H \to e \mu$ must be
 $\lesssim 10^{-5}$~\cite{BEI}. The was some excitement after Run~1 when the combined CMS and
 ATLAS data indicated a possible 2-$\sigma$ excess~\cite{Htaumu}. This has not reappeared
 in early Run~2 data~\cite{CMSHtaumu2}, but remains an open question.
 
 \section{Elementary Higgs Boson, or Composite?}
 
 There has been a long-running theoretical debate whether the Higgs boson could be as
 elementary as the other particles in the SM, or whether it might be composite. The elementary
 option encounters quadratically-divergent loop corrections to the mass of the Higgs boson, 
which are frequently (usually?) postulated to be cancelled by supersymmetric particles~\cite{susy}
appearing at the TeV scale~\cite{hierarchy} - which have not yet been seen. 

On the other hand, the composite
 option has been favoured by many with memories of the (composite) Cooper pairs
 underlying superconductivity, and the (composite) pions associated with quark-antiquark
 condensation in QCD~\cite{techni}. A composite Higgs would require a novel set of strong interactions,
 and early models tended to have a scalar particle much heavier than the Higgs that has
 been discovered, and to be in tension with the precision electroweak data.
  These difficulties can be circumvented by postulating that the Higgs is a pion-like
 pseudo-Nambu-Goldstone boson of a partially-broken larger symmetry that is
 restored at some higher energy scale~\cite{PNGBH}.
 
A phenomenological framework that is convenient for characterizing the experimental
constraints on such as possibility is provided by the following form of effective Lagrangian
that preserves a custodial SU(2)$_V$ symmetry that guarantees $\rho \equiv  m_W/m_Z \cos \theta_W = 1$
up to quantum corrections~\cite{NLEL}:
\begin{eqnarray}
{\cal L} & = & \frac{v^2}{4} {\rm Tr} D_\mu \Sigma D^\mu \Sigma \left(1 + 2 \kappa_V \frac{H}{v} + b \frac{H^2}{v^2} + \dots \right)
- m_i \bar{\psi}^i_L \Sigma \left( \kappa_F \frac{H}{v} + \dots \right) + {\rm h.c.} \nonumber \\
&& + \frac{1}{2} \partial_\mu H \partial^\mu H + \frac{1}{2} H^2 + d_3 \frac{1}{6} \left( \frac{3 m_H^2}{v} \right) H^3
+ d_4 \frac{1}{24} \left( \frac{3 m_H^2}{v} \right) H^4 + ... \, ,
\label{nonlinearH}
\end{eqnarray}
where $H$ is the field of the physical Higgs boson and the massive vector bosons are parametrized by the
$2 \times 2$ matrix $\Sigma = \exp( i \frac{\sigma_a \pi_a}{v} )$. The terms in (\ref{nonlinearH})
are normalized so that the coefficients $\kappa_V, b, \kappa_F, d_i = 1$ in the SM. The question for experiment is
whether any of these coefficients exhibit a deviation that might be a signature of some composite Higgs model.

\begin{figure}[ht]
\begin{center}
\includegraphics[width=5.5cm]{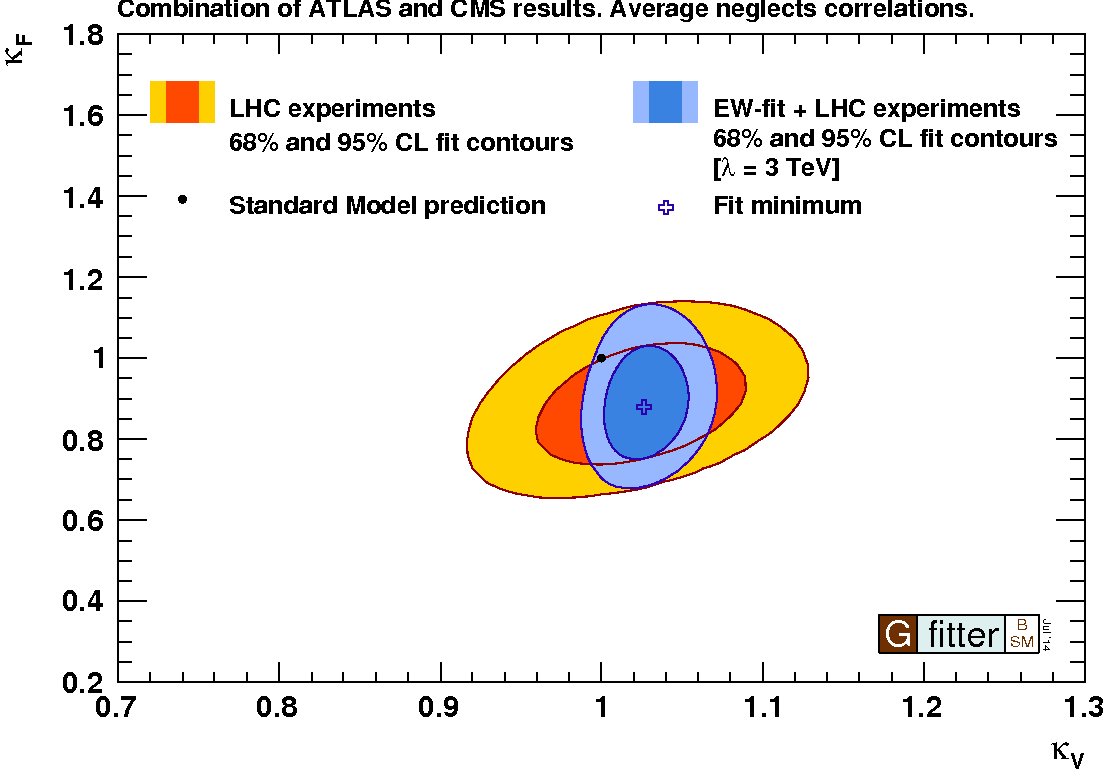}
\includegraphics[width=5.75cm]{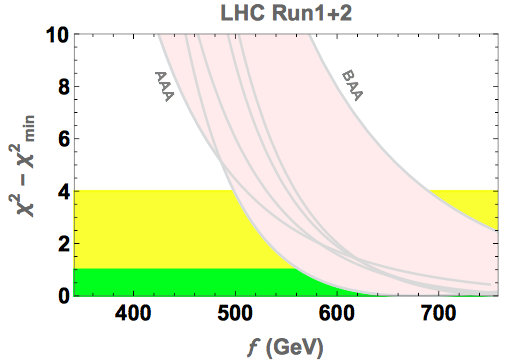}
\caption{\it Left panel: A fit of the LHC $H$ couplings to vector bosons and fermions $(\kappa_V, \kappa_F)$ using $H$ measurements 
 (orange and yellow ellipses), and in combination with precision
electroweak data (blue ellipses)~\protect\cite{Gfitter}. Right panel: Constraints from LHC Run~1 and early Run~2
data on the compositeness scale in various models~\protect\cite{SS}.}
\label{fig:compo}
\end{center}
\end{figure}

As seen in the left panel of Fig.~\ref{fig:compo}, measurements of Higgs properties (yellow and orange ellipses)
and precision electroweak data (blue ellipses) play complementary roles in constraining the $H$ couplings to
vector bosons $\kappa_V$ and fermions $\kappa_F$ in (\ref{nonlinearH})~\cite{Gfitter}. These constraints can be translated into lower limits on the possible 
compositeness scale in various models, as seen in the right panel of Fig.~\ref{fig:compo}~\cite{SS}.

\section{Stability of the Electroweak Vacuum}

\begin{figure}[h!]
\centering
\includegraphics[width=6.1cm]{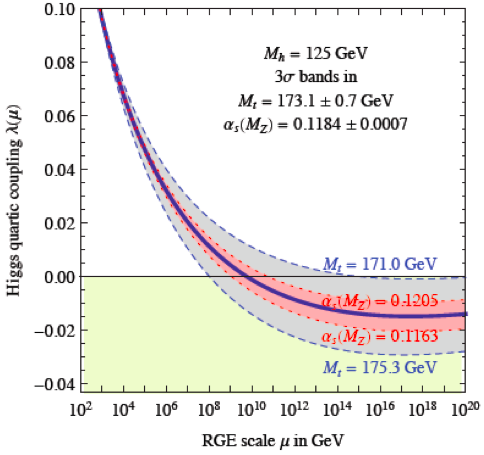}
\includegraphics[width=6cm]{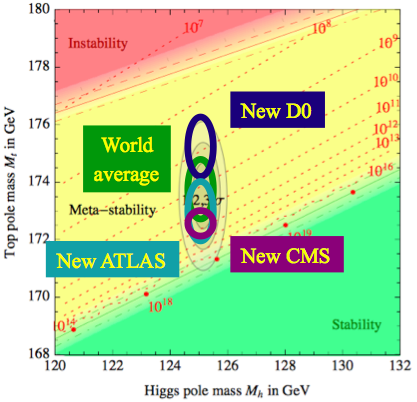}
\caption{\it Left panel: Top quark loops renormalize the Higgs self-coupling $\lambda$ negatively, 
suggesting that it takes negative values at field values $\gtrsim 10^{9}$~GeV\protect\cite{DDEEGIS},
leading to instability of the Higgs potential in the SM.
Right panel: Measurements of $m_t$ and $m_H$ indicate that the SM vacuum is probably
metastable, although there are important uncertainties in $m_t$ and $\alpha_s$~\protect\cite{BDGGSSS}.}
\label{fig:down}
\end{figure}

If the Higgs is indeed elementary, the measurements (\ref{mHRun1}, \ref{mHRun2}) of $m_H$,
combined with those of $m_t$, raise important questions about the stability and history
of the electroweak vacuum, suggesting the necessity of new physics beyond the SM~\cite{Medellin}.
The issue is that the Higgs quartic self-coupling $\lambda$ is renormalized not only by itself,
which tends to increase it as the energy/mass scale increases, but also by the Higgs coupling
to the top quark, which tends to drive it to smaller (even negative) values at higher scales $Q$,
as seen in the left panel of Fig.~\ref{fig:down}GeV~\cite{DDEEGIS}. At leading order:
\begin{equation}
\lambda (Q) \; \simeq \; \lambda (v) - \frac{3 m_t^4}{2 \pi v^4} \log \left( \frac{Q}{v} \right) \, ,
\label{down}
\end{equation}
The right panel of Fig.~\ref{fig:down} displays
the results of one calculation of the regions of the $(m_H, m_t)$ plane where the electroweak
vacuum is stable, metastable or unstable, and yields the following estimate of the `tipping point'
$\Lambda_I$ where $\lambda$ goes negative~\cite{BDGGSSS}:
\begin{eqnarray}
\log_{10} \left( \frac{\Lambda_I}{\rm GeV}\right) & = & 9.4 + 0.7 \left(\frac{m_H}{\rm GeV} -125.15 \right) \nonumber \\
& - & 1.0 \left( \frac{m_t}{\rm GeV} - 173.34 \right) + 0.3 \left( \frac{\alpha_s(m_Z) - 0.1184}{0.0007} \right) \, .
\label{LambdaI}
\end{eqnarray}
The dominant uncertainty in the calculation of $\Lambda_I$ is due to that in $m_t$, followed by that
in $\alpha_s(m_Z)$ (which enters in higher order in the calculation), the uncertainty due to the 
measurement of $m_H$ being relatively small. The final result is an estimate
\begin{equation}
\log_{10} \left( \frac{\Lambda_I}{\rm GeV} \right) \; = 9.4 \pm 1.1 \, ,
\label{LambdaIvalue}
\end{equation}
indicating that we are (probably) doomed, unless some new physics intervenes.

Some people discount this `problem' on the grounds that the prospective lifetime of the
vacuum is much longer than its age. However, there is another issue, namely that fluctuations
in the Higgs field in the very early Universe would have been much larger than now, and would
probably have driven almost everywhere in the Universe into an anti-De Sitter phase
from which there would have been no escape~\cite{aDS}. One could postulate that our piece of the Universe
happened to be extraordinarily lucky and avoid this fate, but it seems more plausible that some
new physics intervenes before the instability scale $\Lambda_I$. Possible such remedies
include higher-dimensional operators in the SM effective field theory (see the next Section),
a non-minimal Higgs coupling to gravity, or a threshold for new physics such as supersymmetry~\cite{ER}
(see later).

\section{The SM Effective Field Theory}

An alternative way of analyzing the Higgs and other data is to assume that all the known particles
(including the Higgs boson) are SM-like, and look for the effects of physics beyond the SM via an
effective field theory (the SMEFT) containing
higher-dimensional SU(2)$\times$U(1)-invariant operators constructed out of SM fields, e.g., of dimension 6~\cite{SMEFT}:
\begin{equation}
{\cal L}_{eff} \; = \; \sum_n \frac{c_n}{\Lambda^2} {\cal O}_n \, ,
\label{dim6}
\end{equation}
where the characteristic scale of new physics is described by $\Lambda$, with the $c_n$ being unknown
dimensionless coefficients. Data on Higgs properties, precision electroweak data, triple-gauge couplings (TGCs), etc.,
can all be combined to constrain the SMEFT operator coefficients in a unified and consistent way. Table~\ref{tab:dim6}
shows which observables currently provide the greatest sensitivities to some of these operators~\cite{ESY}.

\begin{table*}[htb!]
\tbl{
	Some of the relevant CP-even dimension-6 SMEFT operators in the basis.
We display the types of observables that provide the greatest sensitivities to each operator.}	{
	{\begin{tabular}
{|c|c|c|} \hline
EWPTs & Higgs Physics & TGCs \\
\colrule
\multicolumn{3}{| c |}{ ${\mathcal O}_W=\frac{ig}{2}\left( H^\dagger  \sigma^a \lra{D^\mu} H \right)D^\nu  W_{\mu \nu}^a$ } \\
\hline
 & \multicolumn{2}{| c |} {${\mathcal O}_{HW}=i g(D^\mu H)^\dagger\sigma^a(D^\nu H)W^a_{\mu\nu}$ } \\
\hline
 & \multicolumn{2}{| c |}{$ {\mathcal O}_{HB}=i g^\prime(D^\mu H)^\dagger(D^\nu H)B_{\mu\nu}$ } \\
\hline
\multicolumn{1}{| c |}{} & \multicolumn{1}{| c |}{${\mathcal O}_{g}=g_s^2 |H|^2 G_{\mu\nu}^A G^{A\mu\nu}$} & 
\multicolumn{1}{| c |}{${\mathcal O}_{3W}= g \frac{\epsilon_{abc}}{3!} W^{a\, \nu}_{\mu}W^{b}_{\nu\rho}W^{c\, \rho\mu}$} \\
\hline
\multicolumn{1}{| c |}{} & ${\mathcal O}_{\gamma}={g}^{\prime 2} |H|^2 B_{\mu\nu}B^{\mu\nu}$ & \multicolumn{1}{| c |}{} \\
\hline
\end{tabular}}}
	\label{tab:dim6}
\end{table*}

\begin{figure}[ht]
\begin{center}
\includegraphics[width=6.5cm]{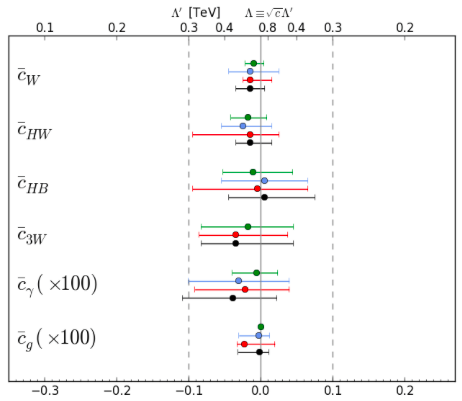}
\includegraphics[width=5.5cm]{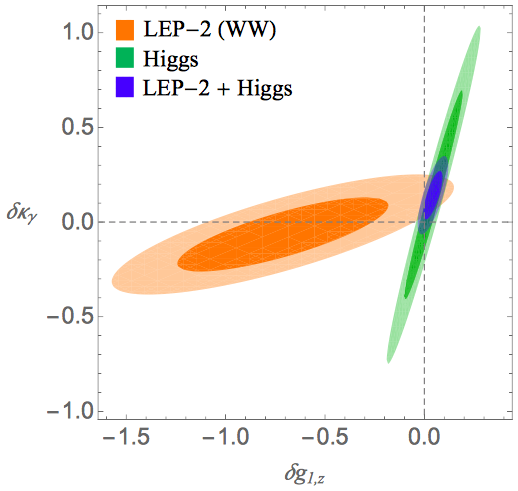}
\caption{\it Left panel: The 95\% CL ranges for fits to individual SMEFT operator coefficients (green bars), 
and the marginalised 95\% CL ranges for global fits combining data on the LHC $H$ signal strength data
with the kinematic distributions for associated $H+V$ production (blue bars), or with the
LHC TGC data (red bars), and combining all the data (black bars)~\protect\cite{ESY}. 
Right panel: The 68 and 95\% CL ranges allowed by a fit to the anomalous TGCs $(\delta g_{1,z}, \delta \kappa_\gamma)$
using LEP-2 TGC data (orange and yellow), LHC Higgs data (green) and their combination (blue)~\protect\cite{Falkowski}.}
\label{fig:SMEFT}
\end{center}
\end{figure}

The left panel of Fig.~\ref{fig:SMEFT} shows how the coefficients of the SMEFT operators in Table~\ref{tab:dim6}
were constrained by Run~1 Higgs data including kinematical variables (blue bar) and by Run~1 measurements
of TGCs (red bar)~\cite{ESY}. The green bar gives the resulting ranges when each operator is switched on individually,
and the black bar is for a global fit marginalizing over all the listed operators. The right panel of Fig.~\ref{fig:SMEFT}
manifests the complementarity between the Higgs and LEP-2 TGC data for constraining the anomalous couplings
$\delta g_{1Z}$ and $\delta g_\gamma$~\cite{Falkowski}. Because of its power to constrain new physics appearing in many
observables in a consistent way, the SMEFT is the preferred framework for assessing the sensitivities of
future analyses of precision LHC measurements to physics beyond the SM, whose motivations are discussed
in the next Section.

\section{The Standard Model is not Enough~\cite{Bond}}

There are many reasons to anticipate the existence of physics beyond the SM, of which I list just 7 here.
1) The prospective instability of the electroweak vacuum discussed earlier. 2) The astrophysical and
cosmological necessity for dark matter. 3) The origin of matter itself, i.e., the cosmological baryon asymmetry.
4) The masses of neutrinos. 5) The naturalness of the hierarchy of mass scales in physics.
6) A mechanism (or replacement) for cosmological inflation to explain the great size and age of the Universe.
7) A quantum theory of gravity.

The good news is that LHC experiments are tackling most of these issues during Run~2. The bad news is
that there is no consensus among theorists how to resolve them. Until recently, supersymmetry found the
most theoretical favour, but the negative results from early Run~2 supersymmetry searches have caused
some to waver. Not me, however - I still think that it is the most comprehensive and promising framework
for new physics beyond the SM. In the words of the famous World War 1 cartoon~\cite{cartoon} "If you knows of a better
'ole, go to it." I do not, so I will stay in the supersymmetric 'ole.

\section{Supersymmetry}

Indeed, I would even argue that Run~2 of the LHC has provided us with 3 new motivations for
supersymmetry. i) It stabilizes the electroweak vacuum~\cite{ER}. ii) It made a successful prediction for the
Higgs mass, namely that it should weigh $\lesssim 130$~GeV in simple models~\cite{susymH}. iii) It predicted
correctly that the Higgs couplings measured at the LHC should be within a few \% of their
SM values~\cite{EHOW}. These new motivations are additional to the classic ones from the naturalness of
the mass hierarchy~\cite{hierarchy}, the availability of a natural dark matter candidate~\cite{EHNOS}, the welcome help of
supersymmetry in making grand unification possible~\cite{susyGUTs}, and its apparent necessity in string theory,
which I regard as the only serious candidate for a quantum theory of gravity.

At this point, I must 'fess up to two pieces of bad news. One is that theorists have also not reached
any consensus on the most promising supersymmetric model, largely because there is no
favoured scenario for supersymmetry breaking. Alternatives range from models in which this is
assumed to be universal at some GUT scale (such as the CMSSM) to models in which all the soft supersymmetry-breaking
parameters are treated entirely phenomenologically as unknown parameters at the electroweak scale (the pMSSM).
The other piece of bad news is that the LHC experiments have found not even a hint of supersymmetry,
despite many searches making different assumptions about the supersymmetric spectrum~\footnote{On
the other hand, they have found no hint of any other physics beyond the SM, despite a similar myriad
of searches.}.

In the following, the negative results of the searches are combined with other measurements to
constrain the parameter spaces of a couple of representative supersymmetric models.

\subsection{Probing a Supersymmetric SU(5) GUT}

The first model we study here is a supersymmetric SU(5) GUT in which the soft supersymmetry-breaking
gaugino masses are assumed to be universal at the GUT scale, whereas the soft supersymmetry-breaking
scalar masses are generation-independent but allowed to be different for the spartners of fermions in the
$\mathbf{\bar{5}}$ and $\mathbf{10}$ representations~\cite{MCSU5}.

\begin{figure}[ht]
\begin{center}
\resizebox{6cm}{!}{\includegraphics{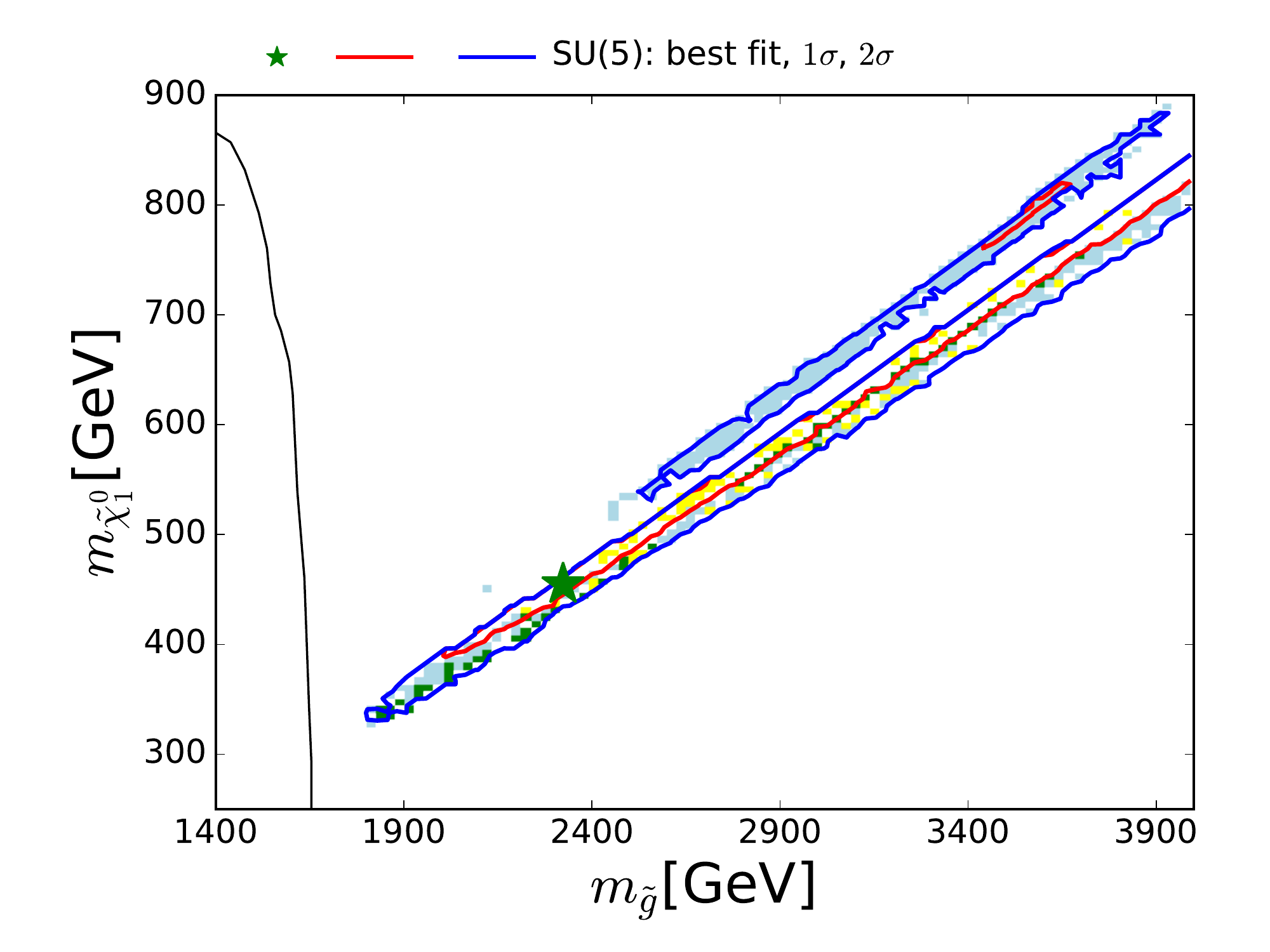}}
\resizebox{6cm}{!}{\includegraphics{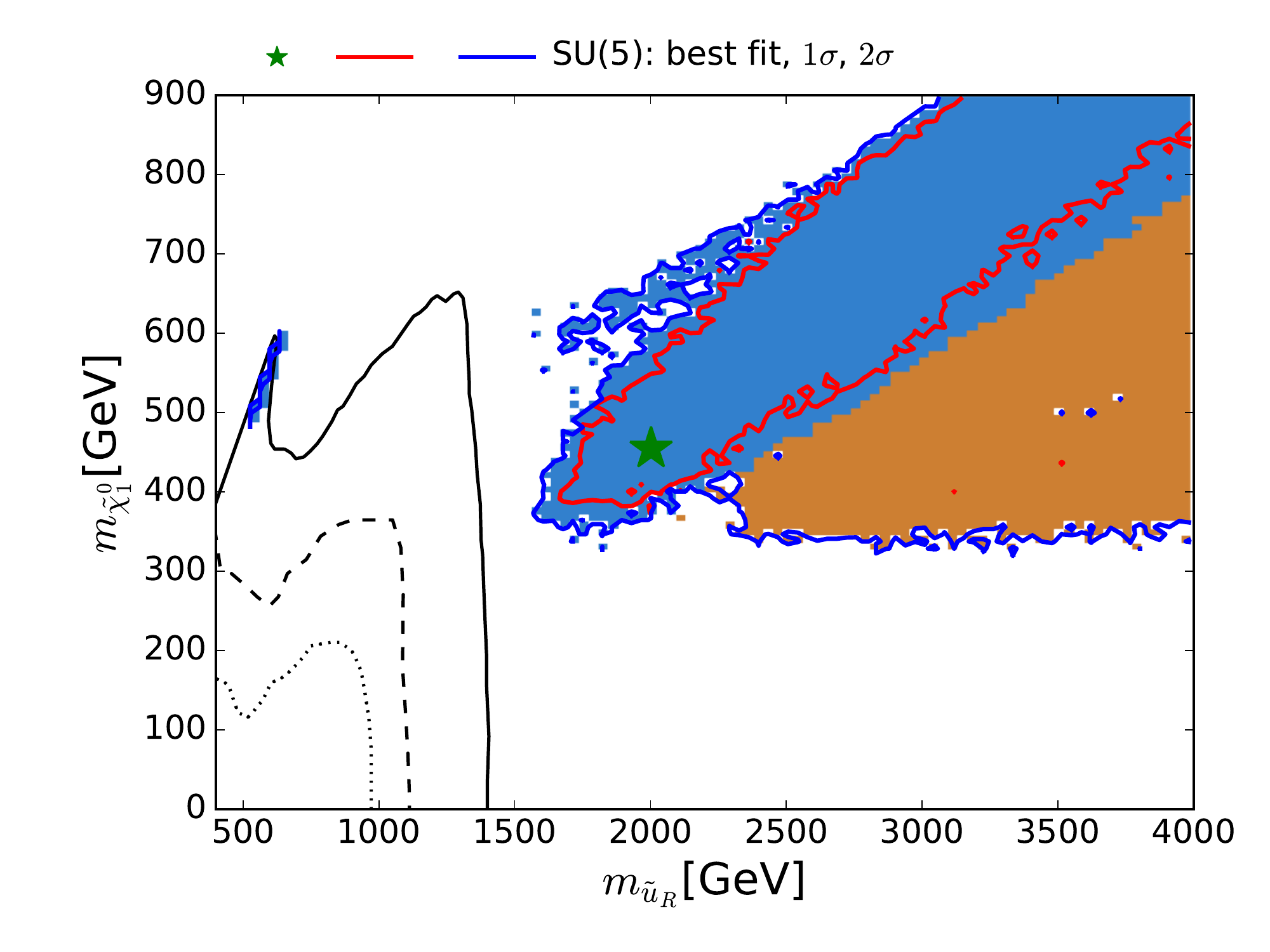}} \\
\resizebox{8cm}{!}{\includegraphics{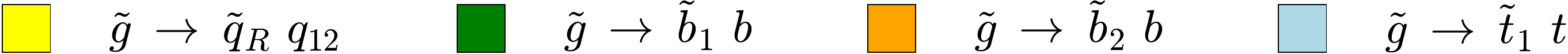}} \hspace{1cm}
\resizebox{3cm}{!}{\includegraphics{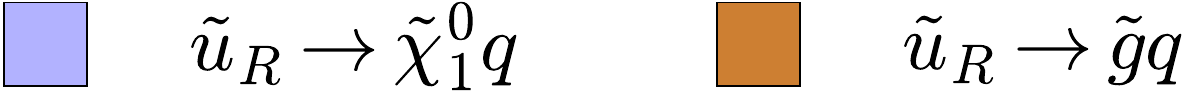}} \\
\caption{\it The 68 and 95\% CL constraints (red and blue contours, respectively) and the best-fit point
(green star) in the $(m_{\tilde g}, m_{\tilde \chi^0_1})$ plane (left panel)
and the $(m_{\tilde u_R}, m_{\chi_1^0})$ plane (right panel) from a global fit in the supersymmetric SU(5)
GUT model~\protect\cite{MCSU5}. The black lines are ATLAS exclusions assuming simplified decay models, whereas the shadings
illustrate the dominant decays in the supersymmetric SU(5) GUT model.}
\label{fig:glsq}
\end{center}
\end{figure}

Fig.~\ref{fig:glsq} displays the regions of the $(m_{\tilde g}, m_{\tilde \chi^0_1})$ plane (left panel)
and the $(m_{\tilde u_R}, m_{\tilde \chi^0_1})$ plane (right panel) that are allowed in a global fit 
in this supersymmetric SU(5) GUT at
the 95\% CL (blue contours) and favoured at the 68\% CL (red contours), as well as the best-fit point
(green stars). The black lines are the nominal 95\% CL limits set by LHC searches, assuming simplified
decay patterns with 100\% branching ratios, and the coloured shadings represent the actual dominant 
decays found in different regions of parameter space. We see in the left panel that gluino masses
$\gtrsim 1900$~GeV are indicated, with a best-fit value of $\simeq 2400$~GeV, whereas the ${\tilde u_R}$
mass may be $\sim 400$~GeV lighter. One curiosity is a small strip in the right panel where
$m_{\tilde u_R} - m_{\tilde \chi^0_1}$ is small and $m_{\tilde u_R} \lesssim 650$~GeV. In this strip
the dark matter (DM) density is brought into the range allowed by astrophysics and cosmology by squark-neutralino
coannihilation, and this compressed-spectrum region is on the verge of exclusion by LHC searches.

\begin{figure*}[htb!]
\begin{center}
\resizebox{8cm}{!}{\includegraphics{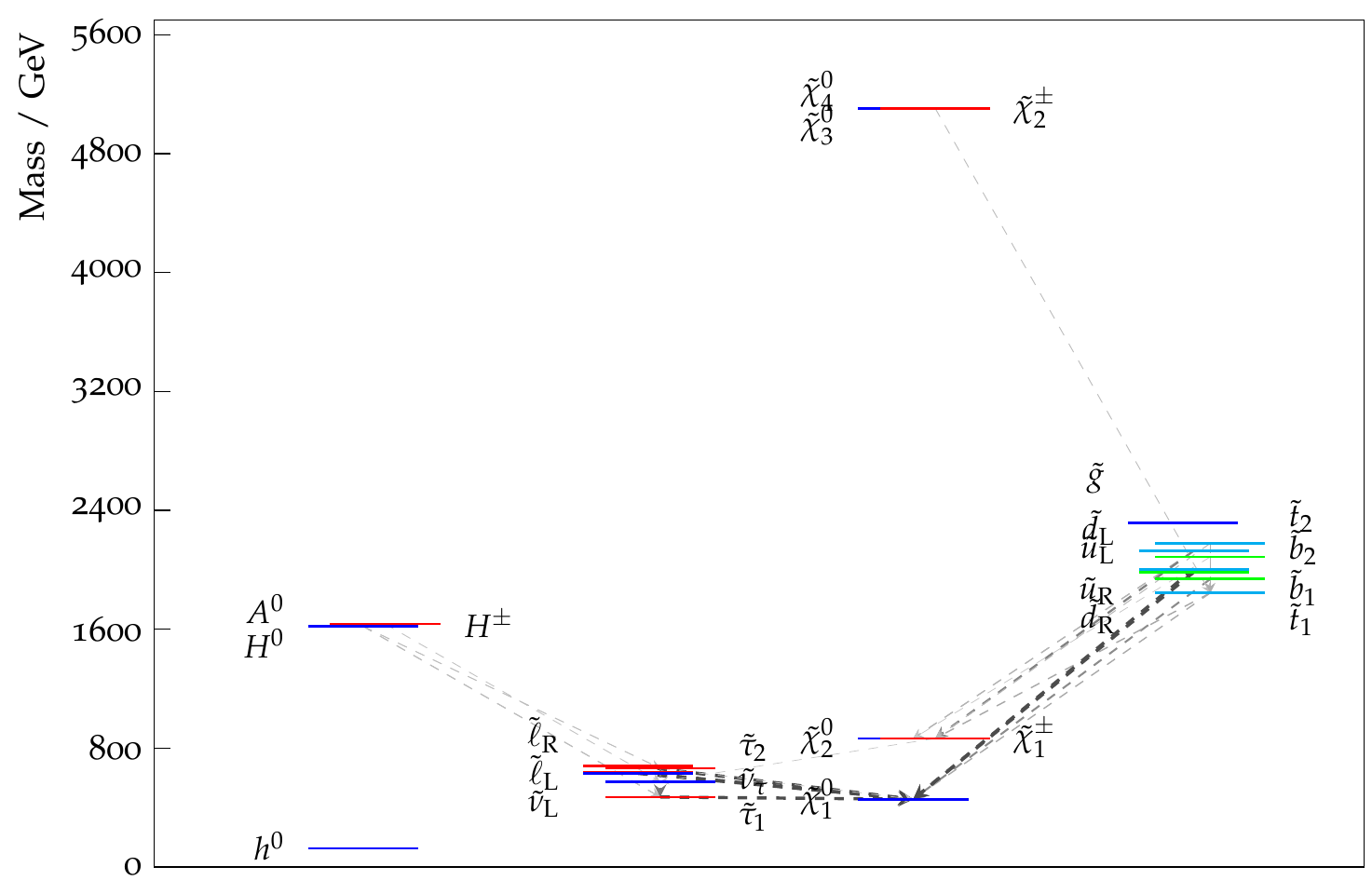}}
\end{center}
\caption{\it 
  The spectrum at the best-fit point in the global fit to the supersymmetric SU(5) GUT model~\protect\cite{MCSU5}.
  The dashed lines indicate decay branching ratios that exceed 20\%.
}
\label{fig:bestfit}
\end{figure*}

\begin{figure*}[htb!]
\begin{center}
\resizebox{6cm}{!}{\includegraphics{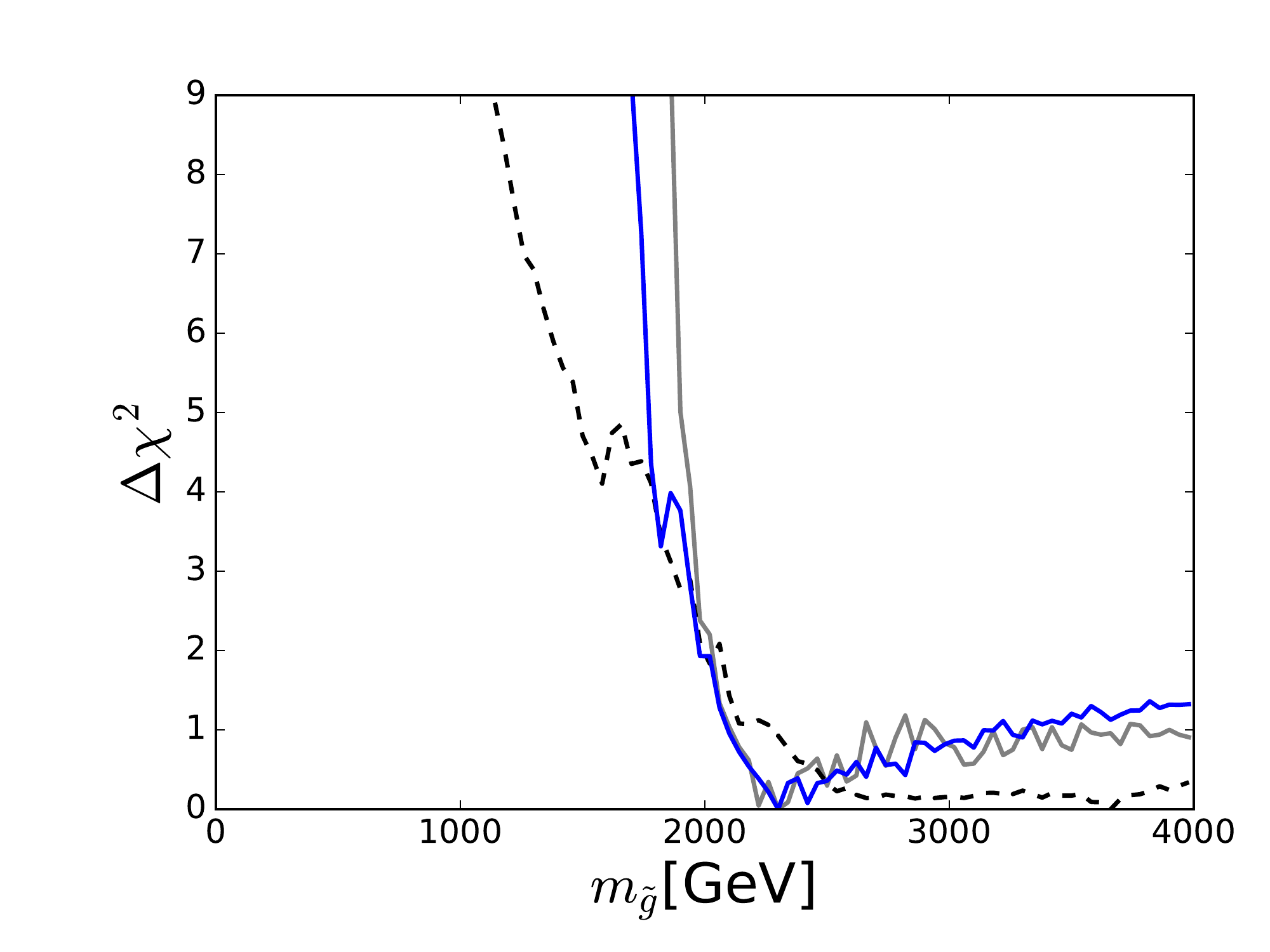}}
\resizebox{6cm}{!}{\includegraphics{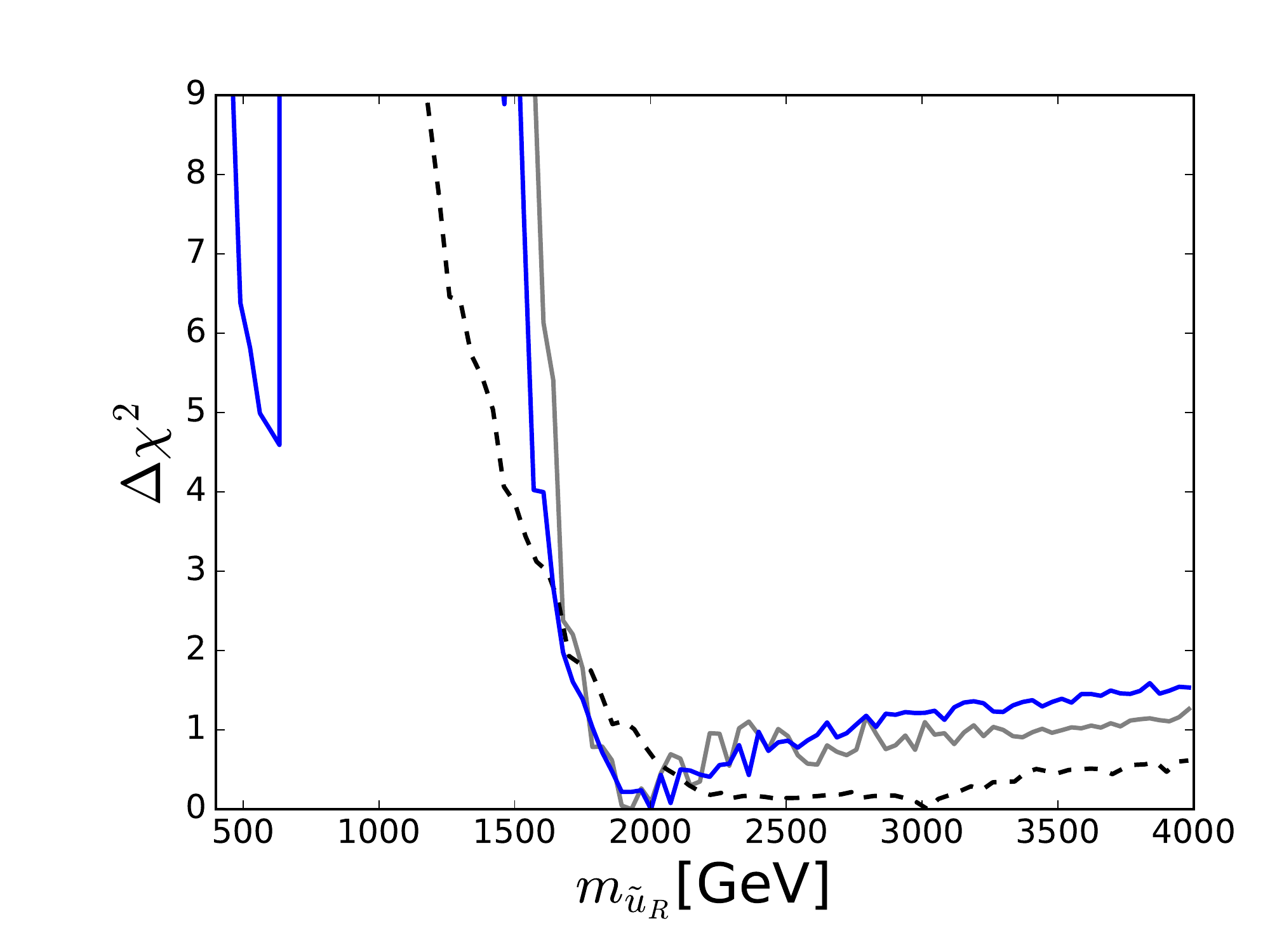}}
\end{center}
\vspace{-0.5cm}
\caption{\it 
The one-dimensional $\chi^2$ likelihood functions for $m_{\tilde g}$ (left panel) and $m_{\tilde u_R}$ (right panel)
in the global fit to the supersymmetric SU(5) model (solid blue lines),
compared with a restriction of the supersymmetric SU(5) GUT model parameters
to emulate a model with universal $\mathbf{\bar{5}}$ and $\mathbf{10}$ scalar masses (solid grey lines)
and a similar fit using only Run~1 LHC data (dashed grey lines), as described in~\protect\cite{MCSU5}.
}
\label{fig:compare}
\end{figure*}

The best-fit spectrum in this SU(5) GUT model is shown in Fig.~\ref{fig:bestfit}. We see that all
the squarks have masses below $\sim 2200$~GeV at the best-fit point, where they would be within the
range of future LHC runs. This analysis included the results from the first $\sim 13$/fb of LHC
data at 13~TeV, and Fig.~\ref{fig:compare} compares the profiled $\chi^2$ likelihood functions for
$m_{\tilde g}$ (left panel) and $m_{\tilde u_R}$ (right panel) found in
this analysis (solid blue lines) with those found in an analysis restricted to 8~TeV data (dashed blue lines)~\footnote{The grey
lines are for the NUHM2 model - see~\cite{MCSU5} for details.}.
We see that, whilst the 13~TeV have had a significant impact, they have not yet been a game-changer.
There is still plenty of room for discovering supersymmetry in future LHC runs in this model, 
though there are no guarantees! 

\begin{figure*}[htb!]
\begin{center}
\resizebox{7cm}{!}{\includegraphics{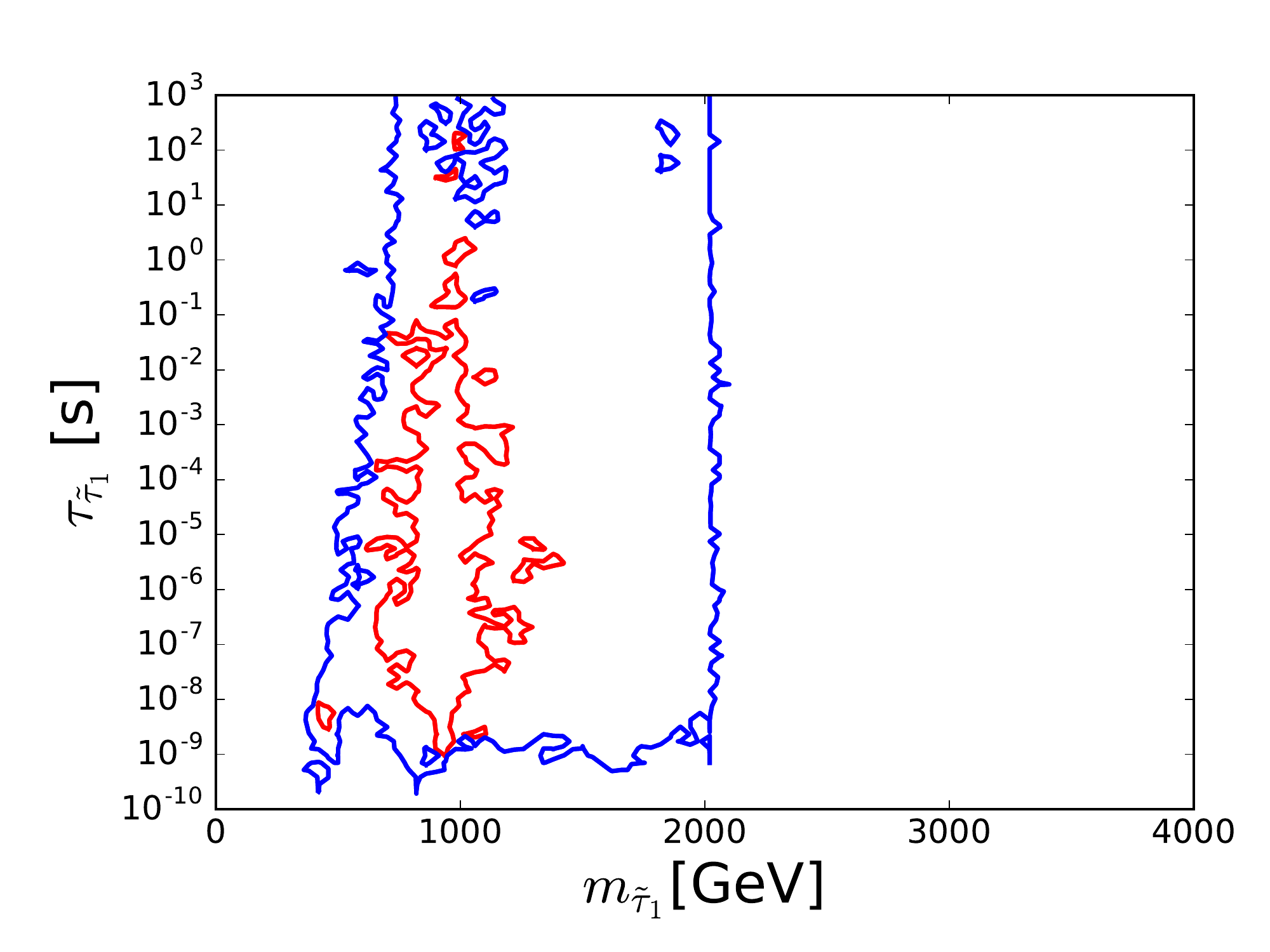}}
\end{center}
\caption{\it 
The $(m_{\tilde \tau_1}, \tau_{\tilde \tau_1})$ plane in the supersymmetric SU(5) model,
showing the 68 and 95\% CL contours (red and blue lines)~\protect\cite{MCSU5}. 
}
\label{fig:staudecay}
\end{figure*}

One of the interesting experimental possibilities in this and related
models is that the next-to-lightest supersymmetric particle (NLSP) might be the lighter stau slepton,
with a mass that could be so close to that of the ${\tilde \chi^0_1}$ that it might have a long enough
lifetime to decay at a separated vertex, or even escape from the detector as a massive charged
non-relativistic particle, as illustrated in Fig.~\ref{fig:staudecay}~\cite{MCSU5}.

\subsection{Probing the Minimal Anomaly-Mediated Supersymmetry-Breaking Model}

\begin{figure*}[htb!]
\begin{center}
\resizebox{12cm}{!}{\includegraphics{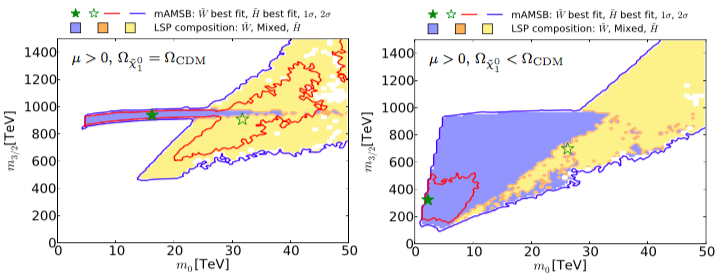}}
\end{center}
\caption{\it 
Planes of the scalar mass $m_0$ and the gravitino mass $m_{3/2}$ in the mAMSB model assuming that the Higgs mixing
parameter $\mu > 0$ and that the LSP provides all the DM (left panel) or only a part (right panel) of the total DM density~\protect\cite{MCmAMSB}. 
The shadings indicate the composition of the LSP.
}
\label{fig:mAMSB}
\end{figure*}

Another model we have studied recently is the minimal anomaly-mediated supersymmetry-breaking
(mAMSB) model~\cite{MCmAMSB}. In this case, the supersymmetric spectrum is relatively heavy. If one assumes that
the lightest supersymmetric particle (LSP) is a wino that provides all the cosmological DM, it must
weigh about 3~TeV, leading to a relatively heavy spectrum as seen in the left panel of Fig.~\ref{fig:mAMSB}, though
the spectrum could be lighter if the LSP is
 a Higgsino, or if it provides only a fraction of the dark matter, as seen in the right panel 
 of Fig.~\ref{fig:mAMSB}. We also see that the
soft supersymmetry-breaking scalar mass $m_0$ in the mAMSB model must be quite large if the
LSP provides all the dark matter, $m_0 \gtrsim 4$~TeV, though it could be smaller if there is some other
contribution to the dark matter.

\begin{figure*}[htb!]
\begin{center}
\resizebox{12cm}{!}{\includegraphics{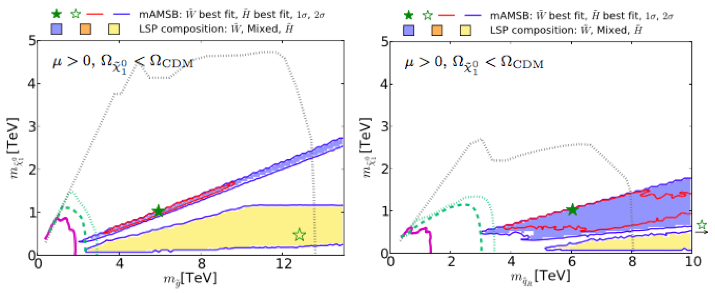}}
\end{center}
\caption{\it 
Planes of $(m_{\tilde g}, m_{\tilde \chi_1^0})$ (left panel) and $(m_{\tilde q_R}, m_{\tilde \chi_1^0})$ (right panel) in the mAMSB model 
for $\mu > 0$ and allowing the LSP to provide only a part of the total DM density~\protect\cite{MCmAMSB}. 
The shadings indicate the composition of the LSP, and the contours indicate the physics reaches of the LHC and of FCC-hh~\protect\cite{FCC-hh-BSM}. 
}
\label{fig:reaches}
\end{figure*}

Fig.~\ref{fig:reaches} displays the reaches of the LHC and a 100-TeV $pp$ collider (FCC-hh) in
the $(m_{\tilde g}, m_{\tilde \chi^0_1})$ plane (left panel) and the $(m_{\tilde q_R}, m_{\tilde \chi^0_1})$ 
plane (right panel) in the mAMSB~\cite{MCmAMSB}. We see that most of the allowed region of the mAMSB parameter space lies
beyond the reach of the LHC, though it may be within reach of FCC-hh~\cite{FCC-hh-BSM}.

\section{Direct Dark Matter Searches}

\begin{figure*}[htb!]
\begin{center}
\resizebox{5.75cm}{!}{\includegraphics{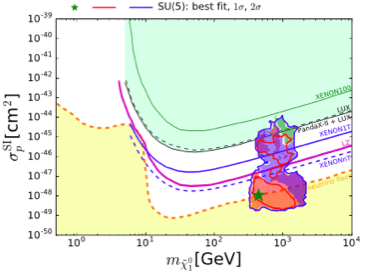}}
\resizebox{6.25cm}{!}{\includegraphics{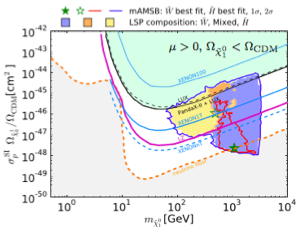}}
\end{center}
\caption{\it 
The $(m_{\tilde \chi_1^0}, \sigma_p^{\rm SI})$ planes in the supersymmetric SU(5) model~\protect\cite{MCSU5} (left panel)
and the mAMSB model~\protect\cite{MCmAMSB} (right panel), indicating the range currently excluded (shaded green),
the sensitivities of planned experiments (blue and purple lines) and the neutrino floor' (dashed orange line)~\protect\cite{DDMAspen}.
}
\label{fig:directDM}
\end{figure*}

Besides missing-energy searches at the LHC, the best prospects for exploring supersymmetry may
be in the direct search for dark matter via scattering on nuclei in deep-underground laboratories~\cite{DDMAspen}.
Possible ranges of the LSP mass and the spin-independent cross section for LSP scattering on
a proton target, $\sigma_p^{\rm SI}$, in the supersymmetric SU(5) and mAMSB models discussed above are shown
in the left~\cite{MCSU5} and right~\cite{MCmAMSB} panels of Fig.~\ref{fig:directDM}, respectively. In both panels 
the range of $\sigma_p^{\rm SI}$ excluded by the latest results from the PandaX~\cite{PandaX} and LUX~\cite{LUX} experiments is shaded green.
The estimated sensitivities of the planned LZ~\cite{LZ} and XENON1/nT~\cite{XENON} experiments are also shown, as is the
neutrino `floor' below which neutrino-induced backgrounds dominate. As in previous plots, the ranges allowed
at the 95\% CL (favoured at the 68\% CL) are surrounded by blue and red contours, respectively, while the
coloured shadings within them correspond to different mechanisms for bringing the LSP density into the
cosmological range (discussed in~\cite{MCSU5,MCmAMSB}, and the best-fit points are marked by green stars.

We see that values of $\sigma_p^{\rm SI}$ anywhere from the present experimental limit down
to below the neutrino `floor' are possible in both the SU(5) and mAMSB cases. There are decent
prospects for discovering direct DM scattering in the LZ and XENON1/nT experiments,
but again no guarantees.

\begin{figure*}[htb!]
\begin{center}
\resizebox{6cm}{!}{\includegraphics{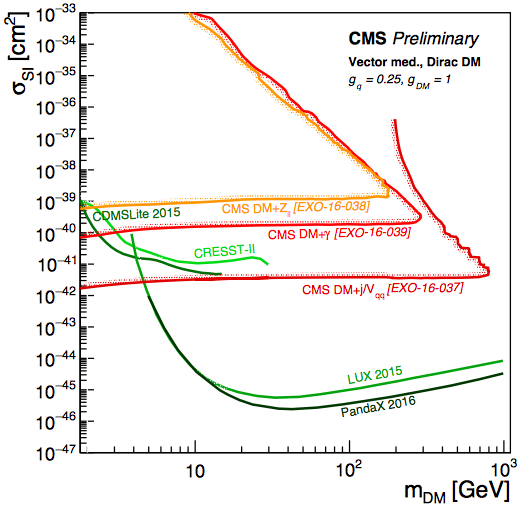}}
\resizebox{6cm}{!}{\includegraphics{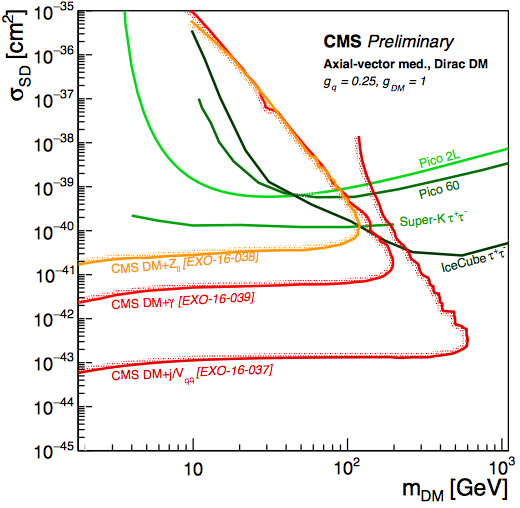}}
\end{center}
\caption{\it 
Comparisons of the LHC and direct DM search sensitivities assuming vector-like couplings (left panel)
and axial-vector couplings (right panel)~\protect\cite{CMSSDMM}. 
}
\label{fig:DMcomparison}
\end{figure*}

It is interesting to compare the sensitivities of LHC searches for mono-jet and other searches with those of direct searches for DM scattering,
which can be done in the frameworks of simplified models for DM~\cite{SDMM}. The results of the comparison
depend, in particular, on the form of the coupling of the intermediate particle mediating the interactions between the DM and SM particles. 
Fig.~\ref{fig:DMcomparison} compares the sensitivities of LHC mono-jet and $\sigma_p^{\rm SI}$
constraints in the case of a vector-like mediator (left panel) and LHC mono-jet searches and constraints
on the spin-dependent scattering cross section, $\sigma_p^{\rm SD}$, in the case of an axial-vector mediator (right panel)~\cite{CMSSDMM}.
We see that in the vector-like case the direct DM searches currently have more sensitivity except for small DM masses,
whereas in the axial-vector case the LHC has greater sensitivity over a wide range of DM masses. These examples
illustrate the complementarity of the LHC and direct searches in the quest for dark matter.

\section{A Plea for Patience}

The LHC will continue to operate for another 15 to 20 years, with the objective of gathering
two orders of magnitude more data than those analyzed so far. Thus it has many opportunities to
discover new physics beyond the Standard Model, e.g., in Higgs studies and in searches for
new particles beyond the Standard Model such as supersymmetry and/or dark matter.
Some lovers of superymmetry may be tempted to lose faith. However, it is worth remembering
that the discovery of the Higgs boson came 48 years after it was postulated, whereas the
first interesting supersymmetric models in four dimensions were written down at the end of 1973~\cite{susy},
only just over 43 years ago! Moreover, the discovery of gravitational waves came just 100 years
after they were predicted. Sometimes one must be patient.

In the mean time, what are the prospects for new accelerators to follow the LHC?

\section{Electron-Positron Colliders}

\begin{figure*}[htb!]
\begin{center}
\resizebox{10cm}{!}{\includegraphics{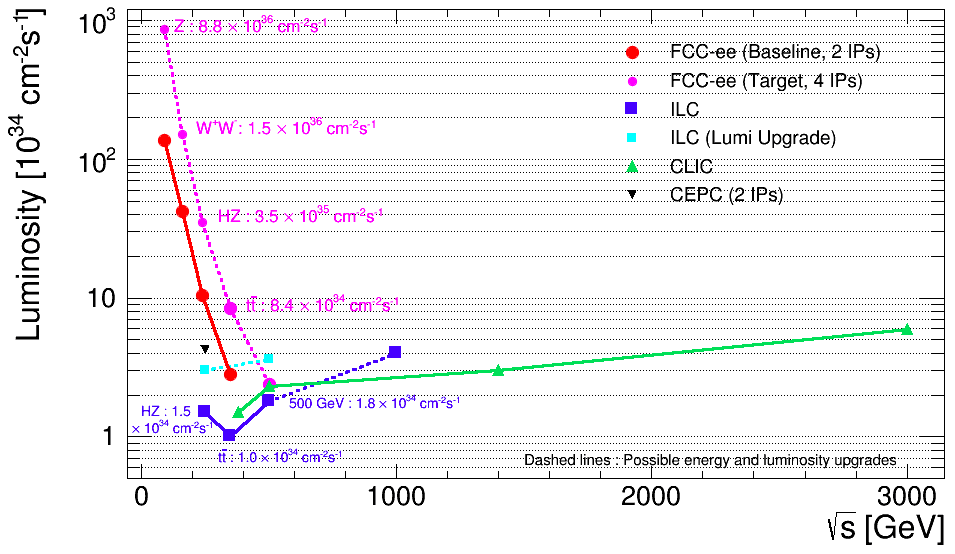}}
\end{center}
\caption{\it 
Comparisons of the centre-of-mass energy reaches of various proposed $e^+ e^-$ colliders
and their design luminosities~\protect\cite{FCC-ee}. 
}
\label{fig:electronpositron}
\end{figure*}

Fig.~\ref{fig:electronpositron} shows the estimated luminosities as functions of the centre-of-mass energy
for various projected $e^+ e^-$ colliders. We see that linear colliders (ILC~\cite{ILC}, CLIC~\cite{CLIC}) could reach higher energies,
but circular colliders (CEPC~\cite{CEPC}, FCC-ee~\cite{FCC-ee}) could provide higher luminosities at low energies. This means that
CLIC, in particular, might be the accelerator of choice if future LHC runs reveal some new particles with
masses $\lesssim 1$~TeV, or if the emphasis will be on probing decoupled new physics via SMEFT
effects that grow with the centre-of-mass energy~\cite{ERSY-CLIC}, whereas FCC-ee would be advantageous~\cite{EY-FCC-ee} if high-precision
Higgs and $Z$ measurements are to be prioritized.

\begin{figure*}[htb!]
\begin{center}
\resizebox{6.5cm}{!}{\includegraphics{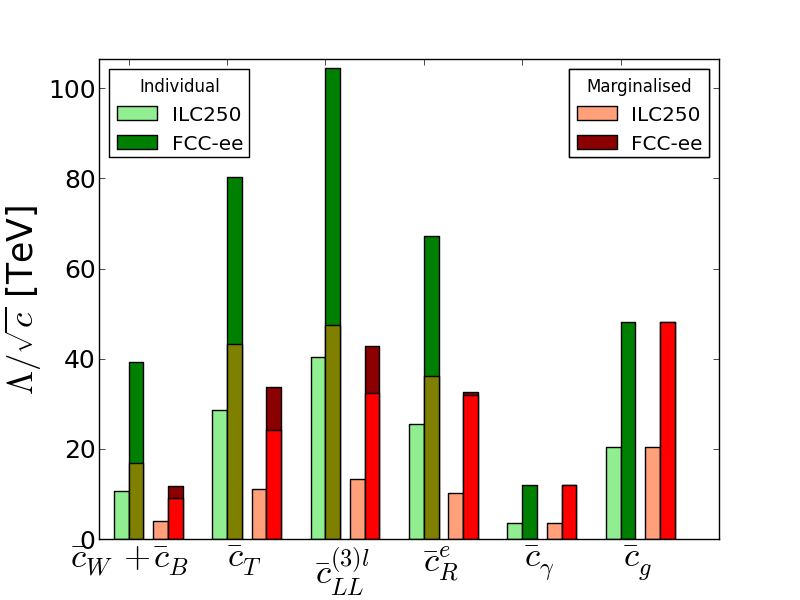}}
\resizebox{5.9cm}{!}{\includegraphics{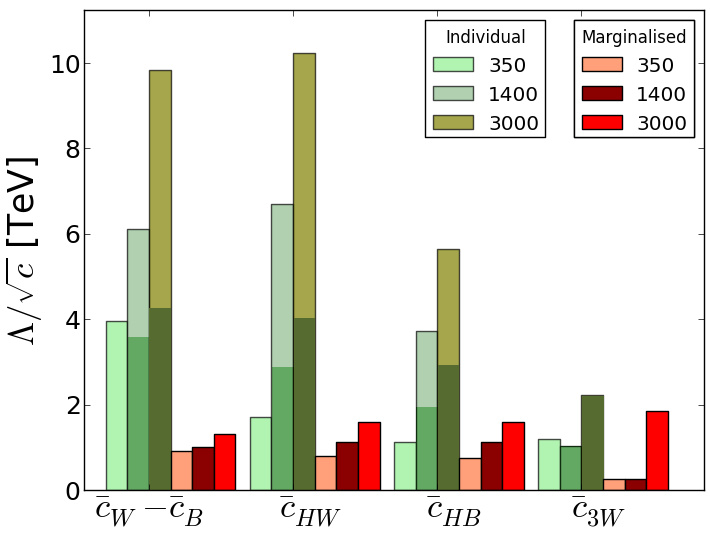}}
\end{center}
\caption{\it 
The sensitivities of possible FCC-ee and ILC measurements to various SMEFT coefficients~\protect\cite{EY-FCC-ee} (left panel),
and of possible CLIC measurements~\protect\cite{ERSY-CLIC} (right panel). 
}
\label{fig:FCC-eeCLIC}
\end{figure*}

The left panel of Fig.~\ref{fig:FCC-eeCLIC} compares the estimated sensitivities of FCC-ee and ILC
measurements of Higgs and electroweak precision measurements to the coefficients of some
dimension-6 operators in the SMEFT~\cite{EY-FCC-ee}. The green bars are for fits to individual operator
coefficients, and the red bars are after marginalization in global fits. We see that both
FCC-ee (darker bars) and ILC (lighter bars) could reach far into the multi-TeV region.
The right panel of Fig.~\ref{fig:FCC-eeCLIC} shows the estimated sensitivities of CLIC
measurements to other combinations of dimension-6 SMEFT operators~\cite{ERSY-CLIC}, highlighting the
advantages conferred by high-energy running at CLIC.

\section{Higher-Energy Proton-Proton Colliders}

Circular colliders with circumferences approaching 100~km are being considered in China (CEPC/SppC~\cite{CEPC})
and as a possible future CERN project (FCC-ee/hh~\cite{FCC}). One could imagine filling the tunnel with two
successive accelerators, as was done with LEP and then the LHC in CERN's present
27-km tunnel.

\begin{figure*}[htb!]
\begin{center}
\resizebox{6.3cm}{!}{\includegraphics{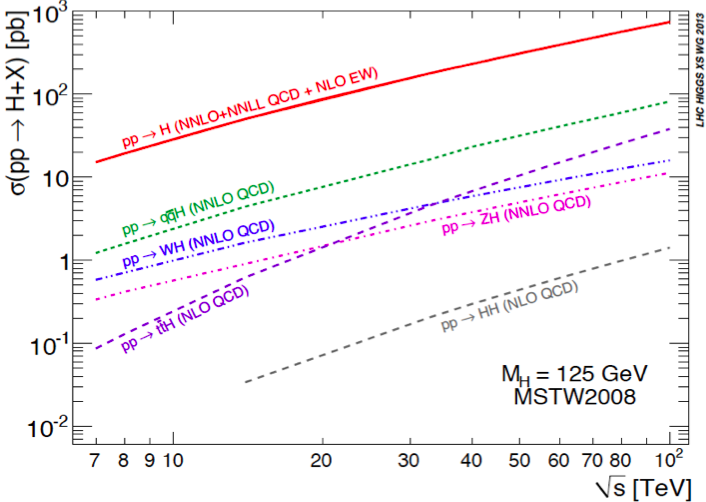}}
\resizebox{5.8cm}{!}{\includegraphics{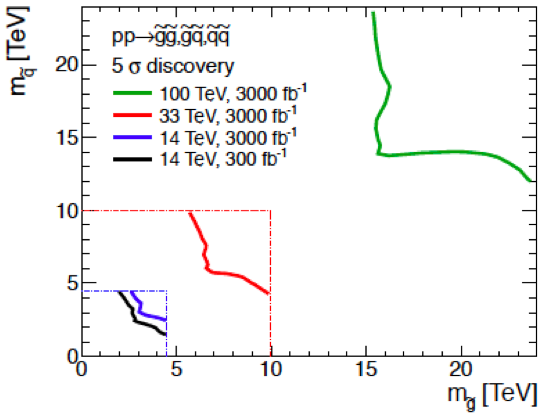}}
\end{center}
\caption{\it Left panel: The energy dependences of the most important Higgs production cross sections in
$pp$ collisions~\protect\cite{FCC-hh-H} (left panel), and the reach of FCC-hh for gluino and squark
production~\protect\cite{FCC-hh-BSM} (right panel).}
\label{fig:FCC-pp}
\end{figure*}

Fig.~\ref{fig:FCC-pp} provides two illustrations of the possible physics reach of the
FCC-hh project for a $pp$ collider. In the left panel we see the ways in which various Higgs production
cross sections grow by almost two orders of magnitude with the centre-of-mass energy~\cite{FCC-hh-H}, offering many possibilities for
high-precision measurements of Higgs production mechanisms and decay modes
in collisions at 100 TeV. In particular, these might offer the opportunity to make the
first accurate direct measurements of the triple-Higgs coupling. In the right panel we
see the discovery reaches for squark and gluino discovery at FCC-pp~\cite{FCC-hh-BSM}. The reaches for
both these sparticles extend beyond 10 TeV and offer, e.g., the prospects for
detecting the heavy spectrum of the mAMSB model shown in Fig.~\ref{fig:reaches}.

In my opinion, the combination of high precision and large kinematic reach offered
by large circular colliders is unbeatable as a vision for the future of high-energy
physics, offer the twin possibilities of exploring the 10~TeV scale
directly in $pp$ collisions at centre-of-mass energies up to 100 TeV and indirectly via the
high-precision $e^+ e^-$ measurements mentioned in the previous Section.

\section{Summary}

Despite the impressive progress already made, many things are still to be learnt about
the Higgs boson, including its expected dominant $b \bar{b}$ decay modes, rare decays
into lighter particles and the triple-Higgs coupling. The best tool for interpreting Higgs
and other electroweak measurements is the SMEFT, and possible future $e^+ e^-$
colliders offer good prospects for higher-precision measurements beyond the sensitivities
of the LHC.

Like that of Mark Twain, rumours of the death of supersymmetry are exaggerated. I still think that it
is the best framework for TeV-scale physics beyond the SM at the TeV scale. 
Simple supersymmetric models have been coming under
increasing pressure from LHC searches, but other models with heavier spectra are
still quite healthy. There are good prospects for discovering supersymmetry in future
LHC runs and in direct dark matter detection experiments, but no guarantees. Maybe
we will have to wait for a future higher-energy $pp$ collider before discovering or
abandoning supersymmetry?

In the mean time, we look forward to whatever indications the full LHC Run 2 date
may provide before choosing what collider we would like to build next,
but the answer to the question in the title may well be ``round in circles".

\section*{Acknowledgments}

The author's work was supported partly by the STFC Grant ST/L000326/1.
He thanks his collaborators on topics discussed here, and thanks the Institute of Advanced Study of the
Hong Kong University of Science and Technology for its kind
hospitality.


\end{document}